\documentclass[%
aip,
amsmath,amssymb,
reprint,
]{revtex4-1}

\usepackage{graphicx}
\usepackage{dcolumn}
\usepackage{bm}

\usepackage[utf8]{inputenc}
\usepackage[T1]{fontenc}
\usepackage{mathptmx}
\usepackage{etoolbox}

\usepackage{blindtext}
\usepackage{graphicx}
\usepackage{tocloft} 
\usepackage{amsmath,mathtools}
\usepackage{float}
\usepackage{amsthm}
\usepackage{physics}
\usepackage{mathtools}
\usepackage{comment}
\usepackage{tabularx}
\usepackage{array}
\usepackage{sepfootnotes}

\usepackage{xcolor}
\usepackage[normalem]{ulem}

\sepfootnotecontent{1st}{APD3 is used for studying the statistical properties of the heralded photonic state in the Klyshko method, and is not used in the conventional calibration method. }
\sepfootnotecontent{2nd}{We note that for the portable, commercial SPDC source in our experiment, there is a small range of pump powers over which we can reliably implement the Klyshko method.  Working within this range, we can ensure that the pump laser is mode-hope free, and with low enough photon generation rates to minimize, and practically avoid, any accidental coincidences from uncorrelated SPDC processes, dark counts, and aferpulsing effects. For this range of pump powers, the experimental data in Figure \ref{fig:de vs dac} shows a near linear behavior, which results in the theoretical model in Eq. (\ref{model equation}), which is a nonlinear model, overfitting the data. As a result, the uncertainties in the estimates $\hat{\eta}^{tot}$ from the fit are smaller than the statistical uncertainties from the data. Therefore, the uncertainty in $\hat{\eta}^{tot}$ is dominated by statistical uncertainties.}\label{footnote 2}

\makeatletter
\def\@email#1#2{%
	\endgroup
	\patchcmd{\titleblock@produce}
	{\frontmatter@RRAPformat}
	{\frontmatter@RRAPformat{\produce@RRAP{*#1\href{mailto:#2}{#2}}}\frontmatter@RRAPformat}
	{}{}
}%
\makeatother
\begin{document}
	
	\preprint{AIP/123-QED}
	
	\title{Effects of multiphoton states in the calibration of single-photon detectors based on a portable bi-photon source}
	\author{S. Pani}
	\affiliation{ 
		Center for Quantum Information and Control, Department of Physics and Astronomy, University of New Mexico, Albuquerque, New Mexico 87131, USA
	}
	
	\author{D. Earl}%
	\affiliation{%
		Qubitekk, Inc., 1216 Liberty Way, Vista, California 92081, USA
	}%
	
	\author{F. E. Becerra$^1$}
	\email{fbecerra@unm.edu}

	\date{\today}
	
	\begin{abstract}
		Single photon detectors (SPDs) are ubiquitous in many protocols for quantum imaging, sensing, and communications. Many of these protocols critically depend on the precise knowledge of their detection efficiency. A method for the calibration of SPDs based on sources of quantum-correlated photon pairs uses single photon detection to generate heralded single photons, which can be used as a standard of radiation at the single photon level. These heralded photons then allow for precise calibration of SPDs in absolute terms.  In this work, we investigate the absolute calibration of avalanche photodiodes (APD) based on a portable, commercial bi-photon source, and investigate the effects of multi-photon events from the spontaneous parametric down conversion (SPDC) process in these sources. We show that the multi-photon character of the bi-photon source, together with system losses, have a significant impact on the achievable accuracy for the calibration of SPDs. However, modeling the expected photon counting statistics from the squeezed vacuum in the SPDC process allows for accurate estimation of the efficiency of SPDs, assuming that the system losses are known. This study provides essential information for the design and optimization of portable bi-photon sources for their application in on-site calibration of  SPDs with high accuracy, without requiring any other reference standard.
	\end{abstract}
	
	\maketitle
	
\maketitle

\section{Introduction} \label{Intro} 

Recent advances in optical quantum technologies have shown significant advantages over their classical counterparts \cite{sidhu2021advances,tse2019quantum,ono2013entanglement,arute2019quantum}. Much of this progress has been enabled by breakthroughs in the efficient generation, manipulation, and detection of quantum states of light in the single photon regime \cite{dello2022advances,chunnilall2014metrology}. Among different technologies for the detection and characterization of quantum light, single photon detectors (SPD)  are essential in diverse applications spanning quantum communications \cite{https://doi.org/10.1049/qtc2.12044,bedington2017progress}, quantum metrology \cite{taylor2016quantum,bhargav2021metrology,kuck2022single}, quantum imaging \cite{madonini2021single}, and photonic quantum computing \cite{slussarenko2019photonic}.

The central role of SPDs in photonic quantum technologies also underscores the need for robust and accurate calibration techniques for reliable implementation of protocols relying on single photon detection  \cite{khomiakova2021investigation,madonini2021single,tsai2021quantum}.
Moreover, devising robust, repeatable, and accurate calibration techniques for SPDs can also facilitate efforts towards developing standards for photonic quantum technologies referenced to some defined primary standard, such as at National Metrology Institutes (NMIs) \cite{tzalenchuk2022expanding}.

Among various performance metrics for SPDs \cite{ware2007calibrating, wayne2017simple, ziarkash2018comparative, tao2020characterize}, detection efficiency is critical in most photonic quantum technologies \cite{etsi2016011, mohageg2022deep}. This fact has driven many efforts to develop techniques for high accuracy calibration of the efficiency of SPDs at the single (or the few) photon level
\cite{chunnilall2014metrology}. The most reliable calibration technique so far is the substitution method, in which the performance of an SPD, the detector under test (DUT), is compared to a reference detector whose calibration is traceable to some reference, such as a national primary standard \cite{dhoska2016improvement,polyakov2007high,cheung2011low,jin2022calibration}. Given its traceability, most of the other calibration techniques are usually evaluated relative to the substitution method \cite{chunnilall2014metrology}.

An alternate approach, first proposed by Klyshko \cite{klyshko1980use}, makes use of the photon correlations from photon pair sources and single-photon heralding detection to prepare heralded single photons. These heralded photons work as an accurate source of radiation to calibrate an SPD. The Klyshko method in principle can be used to determine the efficiency of an SPD in absolute terms.
High accuracy calibration of SPDs based on this method has been realized within national laboratories of standards and metrological institutes \cite{migdall2002intercomparison,ware2007calibrating,polyakov2007high,gerrits2020calibration}, where spontaneous parametric down conversion (SPDC) in nonlinear crystals was used as a source of correlated photon pairs. These demonstrations used specialized equipment and required involved system calibration and characterization, which is difficult to realize outside these metrological institutes. 
Moreover, while the Klyshko method assumes ideal heralded single photons in a single mode, the heralded quantum state from SPDC can significantly deviate from this assumption \cite{hellebek2024characterization}.  All this puts in question the practicality and applicability of this method using SPDC for high accuracy efficiency calibration of SPDs in settings outside specialized metrology laboratories.

\begin{figure}[ht!]
	\centering
	\includegraphics[width=85mm,height=78mm]{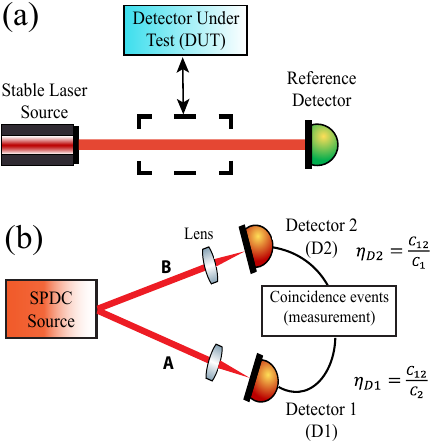}
	\caption{\small{Schematic of (a) the conventional calibration method (the substitution method) and (b) the Klyshko calibration method. The conventional calibration method is implemented by calibrating the input photon flux with a reference standard detector, which is then substituted by the detector under test (DUT)  to determine its detection efficiency. The Klyshko method uses correlated photon pairs to simultaneously calibrate two detectors, D1 and D2. The two photons from each pair are directed to separate single photon detectors. The detection efficiency of the DUTs can be determined in absolute terms by comparing the coincidences $C_{12}$ with the single photon counts from the other detector $C_{1(2)}$ (heralding detector).}}
	\label{fig:conventional schematic}
\end{figure}	
In this work, we investigate the feasibility of the Klyshko method for the absolute calibration of SPDs without requiring traceability to reference standards for its application outside metrological institutes. We employ a commercial portable bi-photon source based on SPDC \cite{qubitekk_entangled_photon_source}, and use coincidence counting to determine the detection efficiency of commercial avalanche photodiode (APD) SPDs in absolute terms.  We theoretically analyze the effects of the underlying properties of the squeezed vacuum state from the SPDC process in the Klyshko method, which ultimately limits the calibration accuracy of SPDs. Finally, we compare the APD detection efficiencies obtained from the Klyshko method with those from the conventional substitution method. This comparison allows us to investigate the effects of non-ideal system conditions on the performance of the Klyshko method.
We observe that multi-photon states from SPDC and the overall system losses have a significant impact on the accuracy of the Klyshko method. This investigation provides an understanding of the required system performances for deployable  SPDC sources for reliable on-site absolute calibration of SPDs with high accuracy.

\section{Background} \label{background} 

The conventional approach for high accuracy calibration of SPDs has mostly relied on comparisons with a transfer standard detector traceable to a primary standard in metrological institutes \cite{migdall2002intercomparison,ware2007calibrating,polyakov2007high,gerrits2020calibration}. A primary standard of optical power, such as a cryogenic radiometer, is used to calibrate the responsivity of the reference detector \cite{biller1999measurements,fox20052,muller2012traceable}, which functions as a secondary standard for calibrating the SPD. Figure \ref{fig:conventional schematic}(a) depicts the main concept of this method. The reference detector measures the mean photon flux from a stable light source with high accuracy. The detection efficiency $\eta^{Conv}_{D}$ of the SPD is then determined by the ratio of the observed count rate ($R_{obs}$) from the SPD to the expected  input photon rate ($R_{exp}$):

\begin{equation}
	\eta^{Conv}_{D}=\frac{R_{obs}}{R_{exp}} .
\end{equation}

Calibration of optical detectors based on the substitution method using sensitive radiometers has achieved very low uncertainties ($\approx0.005\%$) in the high optical power range of 0.1-1 mW \cite{chunnilall2014metrology}. However, this approach cannot yield the same level of precision in the single-photon regime ($\leq1$pW), which is required for calibrating SPDs. This is due to the contributions in uncertainty from the calibration of attenuators to step down the optical power to reach this regime \cite{cheung2011low}. Moreover, since this method relies on a (primary) radiation standard, it makes it difficult to realize high-accuracy calibrations outside national laboratories and metrological institutes.

Alternatively, in the Klyshko method \cite{klyshko1980use},
time-correlated pairs of photons generated at a rate of $N$ pairs/sec, are detected by two detectors D1 and D2 with efficiencies $\eta_{D1}$ and $\eta_{D2}$, respectively, as shown in Figure \ref{fig:conventional schematic}(b).
For ideal time-correlated photon pairs, a photon detection event in one detector, say D1, heralds the existence of a photon in the other mode, which can be detected by D2. Given this information, the detection efficiency of the DUT, say D2 ($\eta_{D2}$), can be determined by observing the coincidence events $C_{12}=\eta_{D1}\eta_{D2}N$ between D1 and D2. The detection efficiency $\eta_{D2}$ of D2 is then given by the ratio of $C_{12}$ to the number of single photon detection events $C_1 = \eta_{D1}N$ observed at the heralding detector D1 during a given measurement time interval: 

\begin{equation}
	\eta_{D2} = \frac{C_{12}}{C_1}=\frac{\eta_{D1}\eta_{D2}N}{\eta_{D1}N}.
	\label{etaD2}
\end{equation}

In a similar way, the detection efficiency $\eta_{D1}$ of detector D1 can be obtained from the ratio of $C_{12}$  to the number of single photon detection events $C_2$ in detector D2: $\eta_{D1} = C_{12}/C_{2}$.

In principle, this method allows for determining the detection efficiency of the DUT in absolute terms, without reference to any standard. Moreover, the accuracy and precision of this method are in principle independent of the efficiency of the heralding detector. These advantages make the Klyshko method a suitable technique for absolute calibration of SPDs and SPD arrays \cite{avella2016absolute} with high accuracy.

However, while this method ideally provides a straightforward path for absolute SPD calibration, it is critical to understand its potential limitations in realistic situations.
Specifically, it is necessary to investigate the achievable accuracy and precision of the method under non-ideal conditions with loss, noise, multi-photon events, and accidental photon counts. We investigate theoretically the effects of multi-photon correlations in the Klyshko method arising from squeezed vacuum states in the SPDC process. Then, we implement this method using a commercial, portable bi-photon source \cite{qubitekk_entangled_photon_source} under realistic conditions of system loss and noise.

\section{Theory}  \label{theory}

The ideal Klyshko method assumes a perfect source of correlated pairs of single photons in two modes.  However, in practice, the experimental realization of this method is done with a photon source based on an SPDC process. In this process, a strong pump propagates through a non-linear crystal producing two correlated output modes that are not necessarily in single-photon states. Moreover, the performance of the Klyshko method is expected to depend on additional systematic effects including photon loss within the nonlinear crystal, channel loss in individual modes, and accidental multi-photon events.

Consider an ideal, lossless type-II SPDC process with two output modes with orthogonal polarizations $\ket{H}$ and $\ket{V}$. Assuming perfect phase matching, no losses, and a single frequency mode, the output from this process is a two-mode squeezed vacuum state \cite{horoshko2019thermal}:
\begin{equation}
	\ket{\psi(r)} = \sech(r)\sum_{n=0}^{\infty}\tanh[n](r)\ket{n}_H\ket{n}_V,
	\label{SqState}
\end{equation}
where $r$ is the squeezing strength, which depends on the pump power and the properties of the nonlinear crystal. $\ket{n}_H$ and $\ket{n}_V$ describe the state with $n$ photons in the horizontal (H) and vertical (V) polarization modes, respectively.

To investigate the Klyshko method based on SPDC, we consider a heralded measurement of the ideal state $\ket{\psi(r)}$  with two detectors: the heralding detector D1, and the detector D2 in the heralded mode. We assume the detectors do not have photon number resolving capability, such as APDs, and have non-unit detection efficiencies. Thus, these detectors have binary outcomes providing click/no-click information. The action of these detectors on the state $\ket{\psi(r)}$ can be described by a positive-operator valued measure (POVM) 
$\{\Pi_{OFF}^{D}, I-\Pi_{OFF}^{D}\}$, where \cite{sempere2022reducing}:
\begin{equation}
	\Pi_{OFF}^{D} = \sum_{n=0}^{\infty}(1-\eta_{D})^n \ket{n}\bra{n}.
	\label{POVMOffDEta}
\end{equation} 

Here $\Pi_{OFF}^D$ and $\Pi_{ON}^{D}=I-\Pi_{OFF}^{D}$ correspond to a zero and a non-zero photon detection event in detector D$i$, respectively, and $\eta_{D}$ describes the single photon detection efficiency of D$i$. 

Assuming a detection event in D1 with detection efficiency $\eta_{D1}$, the conditional (unnormalized) state in the heralded mode B is \cite{horoshko2019thermal}:

\begin{equation}
	\begin{split}
		\tilde{\rho}_{B|D1} &= Tr_{A}\bigl\{\Pi_{ON}^{D1}\ket{\psi(r)}\bra{\psi(r)}\bigr\} \\
		&= \rho_{th}(\zeta) - \frac{1-\zeta}{1-\zeta(1-\eta_{D1})}\rho_{th}(\zeta(1-\eta_{D1})),
	\end{split}	
	\label{thermal difference state} 
\end{equation}
where $Tr_{A}$ means the partial trace over mode A in Fig. \ref{fig:conventional schematic}(b)  and $\rho_{th}(\zeta)$ is the thermal state:

\begin{equation}
	\rho_{th}(\zeta) = (1-\zeta)\sum_{n=0}^{\infty}\zeta^n\ket{n}\bra{n},
	\label{thermal state}
\end{equation}

with $\zeta = tanh^2(r)$ in Eq. (\ref{SqState}) and $Tr(\rho_{th}(\zeta))=1$.

Normalizing the heralded state $\tilde{\rho}_{B|D1}$, we obtain:
\begin{equation}
	\begin{split}
		\rho_{B|D1} &= \frac{\tilde{\rho}_{B|D1}}{Tr(\tilde{\rho}_{B|D1})} \\
		&= \frac{1-\zeta(1-\eta_{D1})}{\zeta\eta_{D1}} \\
		&~~~\times
		\biggl[\rho_{th}(\zeta) - \frac{1-\zeta}{1-\zeta(1-\eta_{D1})}\rho_{th}(\zeta(1-\eta_{D1}))\biggr] .
	\end{split}
	\label{normalized thermal difference state}
\end{equation}

Given the heralded state in Eq. (\ref{normalized thermal difference state}) conditioned on a photon detection in D1, the probability of not observing a photon in detector D2 with detection efficiency $\eta_{D2}$ is:
\begin{equation}
	P_{D2|D1}^{OFF} = Tr\bigl\{\Pi_{OFF}^{D2}\rho_{B|D1}\bigr\},
\end{equation}

where $\Pi_{OFF}^{D2}$ is the measurement operator for zero photon detection in Eq. (\ref{POVMOffDEta}) for D2 with efficiency $\eta_{D2}$. From Eq. (\ref{POVMOffDEta}) and Eq. (\ref{normalized thermal difference state}), $P_{D2|D1}^{OFF}$ becomes:

\begin{multline}
	\begin{split}
		P_{D2|D1}^{OFF} &= \sum_{m=0}^{\infty}\frac{1-\zeta(1-\eta_{D1})}{\zeta\eta_{D1}}(1-\eta_{D2})^m\zeta^m \\
		&\times(1-\zeta)\bigg[1-(1-\eta_{D1})^m\bigg] \\ 
		&= \left(\frac{1-\zeta(1-\eta_{D1})}{\zeta\eta_{D1}}\right)
		(1-\zeta)\\
		&\times\left[\frac{1}{1-\zeta(1-\eta_{D2})}
		-\frac{1}{1-\zeta(1-\eta_{D2})(1-\eta_{D1})}\right].
	\end{split}
	\label{model equation lossless}
\end{multline}
\begin{figure}[t]
	\centering
	\includegraphics[width=85mm,height=66mm]{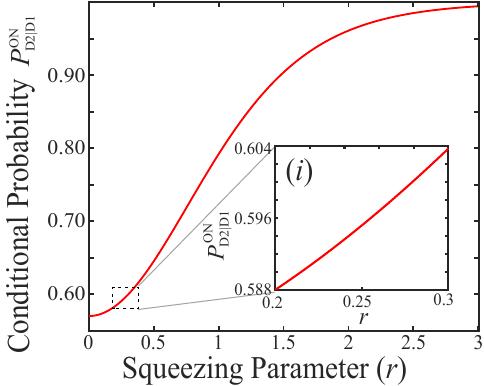 }
	\caption{\small{Conditional probability of detection in Eq. (\ref{model equation}) as a function of squeezing strength $r$ with $\eta_{D1}=0.63$, $\eta_{D2}=0.57$, and $T_{i} = 1$. Inset ($i$) shows a zoom in the region corresponding to our experimental implementation of the Klyshko method with the commercial SPDC source.}}
	\label{fig:theory plot}
\end{figure}
As a result, the probability of a photon detection event in D2 conditioned on having observed a photon in D1 is $P_{D2|D1}^{ON} = 1-P_{D2|D1}^{OFF}$.

This analysis can be carried out considering losses in the SPDC process and optical-channel losses, which unavoidably lead to uncorrelated photons \cite{pereira2013demonstrating,massaro2019improving}. The total losses experienced by the quantum state $\ket{\psi(r)}$ can be described as the propagation through a channel with a non-unit transmittivity $T_{i}$ \cite{jeffers1993quantum} for each mode. Then, this state is incident in a detector with efficiency $\eta_{Di}$. This is equivalent to a situation where an ideal quantum state $\ket{\psi(r)}$ without loss is incident on a detector with overall detection efficiency $\eta^{tot}_{i} = \eta_{Di}T_{i}$ for each mode \cite{hogg2014efficiencies,bonneau2015effect}. Appendix \ref{Klyshko state losses} discusses the analysis for a two-mode squeezed vacuum state $\ket{\psi(r)}$ under loss, characterized by a transmittivity $T_{i}$ for each mode.  Taking this into account, the conditional probability $P_{D2|D1}^{ON}(\eta_{D1},\eta_{D2},r)$ for a quantum state with losses becomes:

\begin{multline}
	P_{D2|D1}^{ON}(\eta^{tot}_{1},\eta^{tot}_{2},r) =1-\biggl\{ \left(\frac{1-\zeta(1-\eta^{tot}_{1})}{\zeta\eta^{tot}_{1}}\right)
	(1-\zeta)\\
	~~\times\left[\frac{1}{1-\zeta(1-\eta^{tot}_{2})}
	-\frac{1}{1-\zeta(1-\eta^{tot}_{2})(1-\eta^{tot}_{1})}\right]\biggr\}.
	\label{model equation}
\end{multline}
	
The conditional probability in Eq.~(\ref{model equation}) ideally corresponds to an absolute measure of the efficiency of detector D2 in the heralded mode, as given in Eq.~(\ref{etaD2}). However, under realistic conditions of non-unit heralding efficiency,  source and channel losses, and multi-photon events from the squeezed vacuum, the conditional probability in Eq.~(\ref{model equation}) significantly deviates from the detection efficiency expected from the idealized Klyshko method.
	
Figure \ref{fig:theory plot} shows the conditional probability of photon detection $P_{D2|D1}^{ON}$ in Eq. (\ref{model equation}) as a function of the squeezing strength $r$ considering no losses in the source and the optical channels $T_{1(2)}=1$, for detection efficiencies $\eta_{D1}=0.63$ and $\eta_{D2}=0.57$ for the heralding D1 and the heralded D2 detector, respectively. These efficiencies correspond to the expected efficiencies in our experiment (see Section \ref{Experiment}). We observe that for low squeezing $r<<1$\, the Klyshko method allows for the accurate determination of the detection efficiency of detector D2, $\eta_{D2}=0.57$. However, its accuracy significantly degrades for larger $r$, saturating at $\eta_{D2}=1$ for $r>>1$. To understand this effect, we note that in the limit of low squeezing ($r\approx0$), the generated state from SPDC in Eq.~(\ref{SqState}) is $\ket{\psi(r)} \approx  \ket{0}_H\ket{0}_V+r\ket{1}_H\ket{1}_V$. Conditioned on a photon detection in D1, the heralded state corresponds to an ideal single photon state $\ket{1}_V$ in the heralded mode. This results in a conditional probability $P_{D2|D1}^{ON}$ close to the expected detection efficiency $\eta_{D2}$ of the DUT D2. On the other hand, as $r$ increases, the probability of generating multi-photon states with higher numbers of photons $n$ from SPDC increases (see Eq.~(\ref{SqState})). These higher order multi-photon states in the heralding arm $\ket{ n\geq 1}_V$ increase the probability of photon detection $P_{D2|D1}^{ON}$, which results in an overestimation of the detection efficiency of the DUT D2.
	
\begin{figure}
	\includegraphics[scale = 1.0]{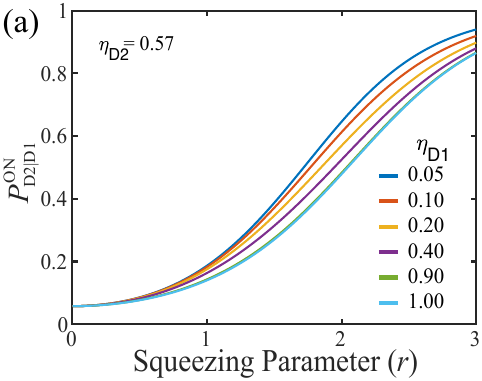}	\includegraphics[scale = 1.0]{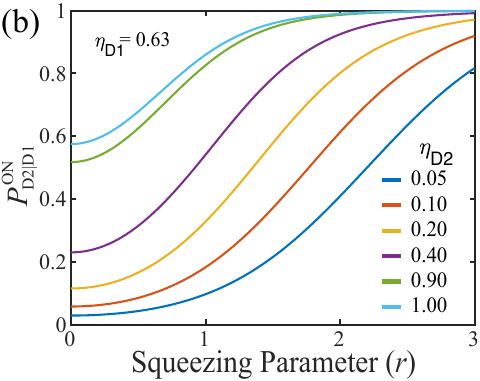}
	\caption{Conditional probability $P_{D2|D1}^{ON}$ as a function of squeezing strength $r$ for  different efficiencies of the heralding $\eta_{D1}$ and heralded $\eta_{D2}$ detectors. (a) Different efficiencies $\eta_{D1}$ for a fixed efficiency $\eta_{D2}$. (b) Different efficiencies $\eta_{D2}$ with fixed efficiency $\eta_{D1}$.}
	\label{fig:Theory_Kly_Eff}
\end{figure}
	
The model in Eq.~(\ref{model equation}) for the conditional probability $P_{D2|D1}^{ON}$ suggests that the optimal regime for the reliable implementation of the Klyshko method based on SPDC is the low squeezing regime ($r\approx0$), where the contributions from multi-photon events are negligible. Beyond this idealized regime, this model provides insight into the accuracy of the Klyshko method for the calibration of SPDs. 
Specifically, the model gives the description of the dependence of $P_{D2|D1}^{ON}$ on the efficiencies of the heralding ($\eta_{D1}$) and heralded ($\eta_{D2}$) detectors, as shown in Fig.~\ref{fig:Theory_Kly_Eff}. Moreover, this model can be used to accurately determine the efficiencies of the SPDs based on the Klyshko method beyond the idealized regime $r\approx0$, given a sufficiently large dynamic range for $r$, and known or minimal source and channel losses $T_{i}\approx$1. Specifically, the efficiencies $\eta_{Di}$ can be obtained from simultaneous fits to two curves $P_{D2|D1}^{ON}(\eta^{tot}_{1},\eta^{tot}_{2},r)$ and $P_{D1|D2}^{ON}(\eta^{tot}_{2},\eta^{tot}_{1},r)$ as a function of $r$, which can be controlled by the pump power.
	
\begin{figure*}[t]
	\centering
	\includegraphics[scale=1.055]{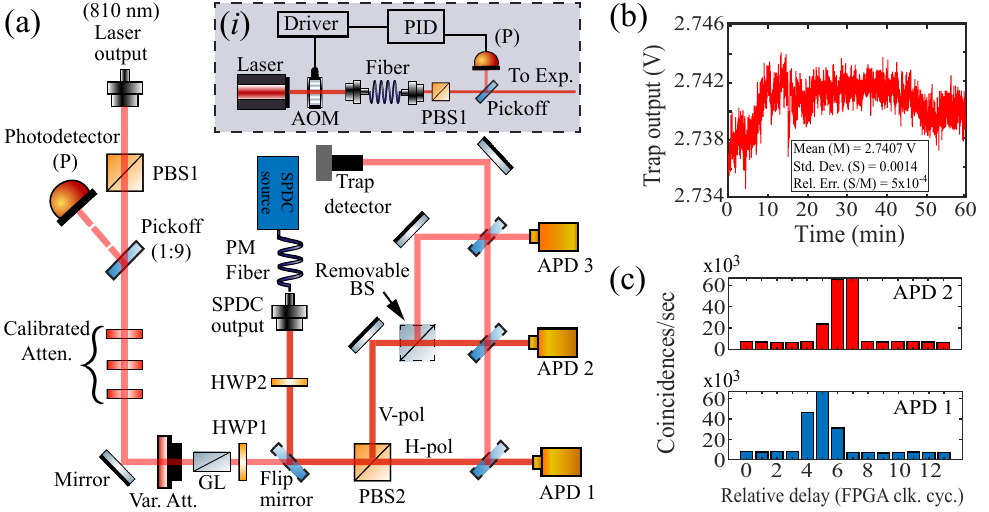}
	\caption{\small{Experiment. (a) Schematic of the experimental setup for implementing the conventional and the Klyshko calibration methods. Inset ($i$) Laser intensity stabilization. (b) Intensity stability of the locked laser as a function of time. (c) Histogram of the coincidence counting statistics for APD1 and APD2 as a function of relative delay in units of clock cycles defined by the (100 MHz) master clock of the FPGA.} APD3 is used to investigate the statistical properties of the heralded state in the Klyshko method. However, this detector is not used during the APD calibration process.}
	\label{fig:Setup Diagram}
\end{figure*}
	
\section{Experiment} \label{Experiment}
	
Figure \ref{fig:Setup Diagram} shows the experimental configuration for the demonstration of the Klyshko method for detector calibration based on a commercial, portable bi-photon SPDC source \cite{qubitekk_entangled_photon_source}. The setup also allows for realizing SPD calibration based on a transfer standard trap detector. Our experiment enables the comparison between these methods under nearly identical experimental conditions including background, noise, temperature, humidity, etc.,  reducing the potential biases and systematic errors. First, we implement the conventional calibration with the transfer standard detector, which provides a benchmark for state-of-the-art SPD calibration. Subsequently, we implement the Klyshko method with the commercial SPDC source.	

\subsection{Calibration with transfer standard detector}
 For the conventional calibration method, we use an intensity stabilized laser diode at 810 nm with 2 nm linewidth. The first diffraction order of an acousto-optic modulator (AOM) is coupled into a single mode fiber followed by a polarizer, as shown in Fig. \ref{fig:Setup Diagram}(a) and inset ($i$). A pick-off directs light to a fast photodetector to monitor the instantaneous intensity of the laser. This signal is used to feed back to the AOM to stabilize the laser intensity to better than 5:10000. Figure \ref{fig:Setup Diagram}(b) shows the intensity variation for the stabilized laser over an hour, as observed with a low noise trap detector. This trap detector is also the transfer standard detector used for absolute power measurements to perform the standard calibration method. The calibration of the efficiency and spectral responsivity of this detector can be traced back to NIST through a secondary standard detector with very low uncertainty (0.05\%)  \cite{gentec_trap7, eppeldauer2009optical}. The stabilized laser source provides a highly stable photon flux. We use this source, together with the trap detector, to calibrate a series of absorptive attenuators to attenuate the light to the few photon level with high accuracy, which is required for calibrating the APDs. For our experiment, the attenuators are calibrated to provide a stable, highly accurate, and controllable photon flux ranging from $10^5-10^6$ photons/second. After the attenuators, 
the light propagates through a high extinction ratio ($\approx10^6$) Glan Laser (GL) polarizer and is then directed to the APD to be calibrated using a polarizing beam splitter (PBS2) and a half-wave plate (HWP1).
In our experiment, we perform the calibration of two commercial APDs: APD1 (SPCM-AQR-13, Perkin Elmer) and APD2 (COUNT-100C, Laser Components \sepfootnote{1st} ), which are calibrated independently.
To calibrate the photon flux incident in the specific APD being calibrated, we first direct the light to the trap detector with a removable, flip mirror. This calibration, together with the calibration of the attenuators with the trap detector, provides a highly accurate estimate of the expected photon flux. The light is then directed back to the APD to be calibrated.
The output from the APD is collected and processed by a field programmable gate array (FPGA). Finally, the measured photon count rates are compared to the expected photon flux to estimate the detection efficiency of the APD under test.

\subsection{Klyshko method}
 In our experimental demonstration of the Klyshko method \cite{klyshko1980use,polyakov2007high}, the commercial bi-photon source is based on a Type-II SPDC process in a PPKTP crystal \cite{qubitekk_entangled_photon_source}. A 405 nm pump beam with adjustable power generates correlated photon pairs in free space, signal and idler, with orthogonal polarizations, H and V, respectively, at 810 nm.
Inside the source module, the correlated photons are coupled into a polarization maintaining (PM) fiber, with associated coupling losses due to mode mismatch. A PM fiber directs the output of the source to the experimental setup, and the correlated photons are launched into free space in a well-defined single spatial mode using a low-loss aspheric lens. The polarizing beamsplitter PBS2 separates the photons into the two polarization modes, which are then directed and focused into the APDs to be calibrated based on the Klyshko method \cite{klyshko1980use,polyakov2007high}. In this method, either one of the detectors can act as the trigger detector (say APD1). A photon detection event on this detector ideally heralds the existence of a photon incident on the other detector to be calibrated (say APD2).  The Klyshko method provides an estimate of the detection efficiency of the APDs based on coincidence counting and single photon rates, as described in Section \ref{background}.

The reliable implementation of the Klyshko method with finite-bandwidth SPDs and photon-counting electronics requires the determination of the optimal coincidence detection window and relative delay between the detector outputs. The optimal coincidence window should capture all the coincidences from the SPDC process while avoiding accidental coincidences due to background, dark counts, and uncorrelated photons. This optimal coincidence window avoids any under or over estimation of the coincidences, which would result in the under or over estimation of the detection efficiency.

Figure. \ref{fig:Setup Diagram}(c) shows the coincidences between the two detectors with a temporal resolution defined by the FPGA clock (10 ns in this experiment), which allows us to determine the optimal relative delay and coincidence window as 10 ns and 30 ns, respectively. The accidental coincidences observed at long relative delays, away from the optimal delays, are subtracted from the observed coincidences to determine the effective coincidence rates from correlated photon pairs (see Appendix \ref{accidental appendix} for details about the characterization of accidental coincidences in our experiment). The ratio of the effective coincidence count rate to the single photon detection count rate at the trigger (heralding) detector provides a direct measurement of the detection efficiency of the DUT. We note that the relatively large background coincidences in our experiment are mainly due to the finite bandwidth of the photon-counting electronics and the relatively large coincidence window (30 ns). However, an accurate characterization of these accidental coincidences and accounting for their contribution to the observed photon-counting statistics avoids any biases in the calibration results.
	
\subsection{Effects of multi-photon states: $\mathbf{g^{(2)}(0)}$ measurement} 
The Klyshko method assumes an ideal source of heralded single photons for the determination of the SPD detection efficiency. However, as discussed in Sec. \ref{theory}, the bi-photon source based on SPDC generates a squeezed vacuum, which is a multiphoton quantum state. As a result, the heralded photonic state unavoidably contains multi-photon components, and the assumption of an idealized source of heralded single photons introduces biases in the estimation of detection efficiency. To evaluate the applicability of the Klyshko method with an SPDC source, we investigate the single-photon character of the heralded photonic mode by studying the conditional autocorrelation function at zero delay $g^{(2)}(0)$ \cite{beck2007comparing}. For an ideal single photon state, $g^{(2)}(0)=0$, while for a Poissonian, uncorrelated photon distribution $g^{(2)}(0)=1$.
	
For this study, APD1 acts as the trigger detector, which, given a photon detection, ideally heralds the existence of a single photon in the vertical polarization after PBS2 (see. Fig. \ref{fig:Setup Diagram}(a)). The heralded field is further split into two modes, followed by single photon detection in each mode by APD2 and APD3. For a perfect heralded single photon, each detection event at APD1 should yield a single detection event either at APD2 or APD3. For the non-ideal case in which the heralded state is a multi-photon state, this would yield non-zero coincidences between APD2 and APD3. Therefore, the $g^{(2)}(0)$ measurement allows for investigating the single-photon character of the heralded field \cite{stasi2023high}. 
	
\begin{figure}[h]
		\centering
		\includegraphics[width=86mm,height=66mm]{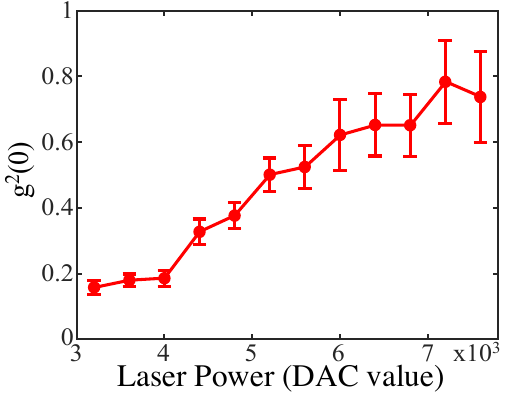}
		\caption{\small{Measurement of the heralded $g^2(0)$ autocorrelation function for the heralded photonic state as a function of SPDC pump DAC value, which corresponds to a digital (near-linear) scale for the pump power.}}
		\label{fig:g2 plot}
	\end{figure}

Figure \ref{fig:g2 plot} shows the measured $g^{(2)}(0)$ as a function of the pump power in the SPDC source (in units of DAC values) for the optimal detection window and relative delays between the heralding detector APD1 and detectors APD2 and APD3. The pump power is adjusted through a digital-to-analog converter (DAC), corresponding to the digital, near-linear scaling for the optical power. 
We note that there is not a perfect linear relation between the DAC values and the pump power throughout the full range of possible powers. To circumvent any potential issue, we determine the optimal operating regime for the pump power by choosing a region where the coincidence rate is much higher than the background and accidental counts, while avoiding regions where sudden jumps in the single photon rates are observed, most likely due to mode hopes in the pump laser. Within the optimal operating range of pump powers for this SPDC source, the minimum observed $g^{(2)}(0)$ is close to $0.16$. We note that while this value is higher than the ideal case of $g^{(2)}(0)\approx0$ \cite{stasi2023high},  $g^{(2)}(0)\approx0.16$ corresponds to a single-photon heralded fidelity of $F_{h}=92\%$ \cite{meyer2020single}, which is consistent with standard SPDC sources \cite{meyer2020single,kaneda2016heralded,christ2012limits}. We further note that in general, ideal photon-pair SPDC sources will yield  $g^{(2) }(0)>0$ for any $r>0$. Therefore, heralding true single photons would require the ability to resolve photon numbers  \cite{christ2012limits}.
Moreover, there are other experimental effects that contribute to the increase in $g^{(2)}(0)$ including crystal fluorescence, background, afterpulsing, etc. These detrimental effects can be overcome with appropriate technical upgrades and optimization of  SPDC sources and other devices.  
		
\section{Results and Discussion} \label{Results}

We implemented the conventional calibration method, based on the transfer standard detector, and determined a detection efficiency of $\eta_{D1}^{Conv}=0.638(3)$ for APD1 and $\eta_{D2}^{Conv}=0.575(2)$ for APD2. The total uncertainties in these measurements account for the uncertainty from the trap detector, the uncertainty from the calibration of the attenuators, and the statistical uncertainties from photon counting measurements, (see Appendix \ref{AppendixLoss}, Table \ref{Conv Uncertainty}). We note that the uncertainties obtained in our experiment are comparable with the typical uncertainties achieved within metrology institutes for the calibration of SPDs based on the conventional methods \cite{polyakov2007high}. We use the results from the conventional method as a benchmark to investigate the performance of the Klyshko method to calibrate the same SPDs. 
		
\begin{figure}[h]
	\centering
	\includegraphics[scale = 1.1]{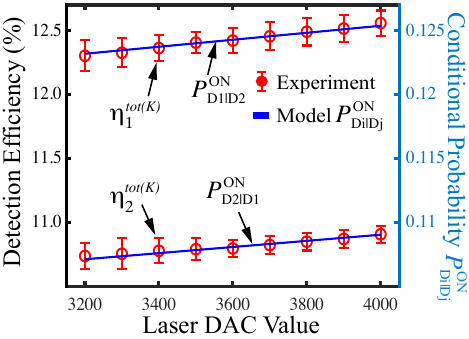}
	\caption{\small{Detection efficiencies from the ideal Klyshko method from Eq. (\ref{etaD2}) and conditional probabilities from Eq. (\ref{model equation}), as a function of pump power (in DAC values) of the SPDC bi-photon source. Red circles show the experimental data with error bars corresponding to 1-$\sigma$ standard deviations from photon counting. Blue curves are the fitted model from Eq. (\ref{model equation}), with  $P_{D2|D1}^{ON}(\eta^{tot}_{1},\eta^{tot}_{2},r)$ and  $P_{D1|D2}^{ON}(\eta^{tot}_{2},\eta^{tot}_{1},r)$ as the model for APD1 and APD2, respectively. $r=\mu\sqrt{KP}$, where $\mu = 0.115$ for the PPKTP crystal \cite{kaiser2016fully}, and $K$ is a proportionality factor to convert DAC values to optical power $P$}.}
	\label{fig:de vs dac}
\end{figure}
		
Figure \ref{fig:de vs dac} shows the detection efficiencies for APD1 and APD2 determined using the Klyshko method as a function of the laser pump power in the SPDC source. The red markers are the estimated efficiencies from the ideal Klyshko method using Eq. (\ref{etaD2}) calculated from the ratios of coincidences and single photon counting rates, with error bars indicating 1-$\sigma$ statistical uncertainty. The two blue curves are fits to the model in Eq. (\ref{model equation}) for  $P_{D2|D1}^{ON}(\eta^{tot}_{1},\eta^{tot}_{2},r)$ and  $P_{D1|D2}^{ON}(\eta^{tot}_{2},\eta^{tot}_{1},r)$, where D1 and D2 correspond to APD1 and APD2, respectively. The fit parameters correspond to the total detection efficiencies $\eta^{tot}_{1(2)}$ and the squeezing strength $r=\mu\sqrt{KP}$, where $K$ is the fit parameter, $P$ is the pump power in DAC values, and $\mu = 0.115$ for the PPKTP crystal \cite{kaiser2016fully}. The models $P_{D2|D1}^{ON}(\eta^{tot}_{1},\eta^{tot}_{2},r)$ and $P_{D1|D2}^{ON}(\eta^{tot}_{2},\eta^{tot}_{1},r)$ are fitted simultaneously to estimate the total efficiencies $\eta^{tot}_{1}$ and $\eta^{tot}_{2}$. From this fit, we infer that the operating regime of pump powers for the source corresponds to squeezing strengths $r\approx0.2-0.3$ (see inset (\textit{i}) in Fig. 2), which results in a near-linear dependence of $P_{D2|D1}^{ON}$ with pump power.

We observe that the total efficiencies determined from the ideal Klyshko method $\eta^{tot (K)}_{i}$ based on Eq. (\ref{etaD2}) increase with pump power. This effect is clearly explained by the theoretical model in Eq.~(\ref{model equation}), which predicts an increase in detection probability with $r$, due to increased multi-photon events. Moreover, we observe that the total efficiencies from this method are much smaller than the APDs efficiencies determined from the conventional calibration method using a transfer standard detector $\eta_{Di}^{Conv}$. This effect is due to losses in the SPDC source together with losses in the optical channel, given that $\eta^{tot}_{i} = \eta_{i}T_{i}$.

We can use the model in Eq.~(\ref{model equation}) to obtain a more accurate estimation of the total detection efficiencies $\eta^{tot}_{i}$. From the fit to the model, we obtain the estimated total efficiencies to be $\eta^{tot (KM)}_{1}=0.114$  and $\eta^{tot (KM)}_{2}=0.099$, 
with an uncertainties $\sigma_{\eta_{i}^{tot}}\approx1\times10^{-3}$, mostly dominated by the statistical uncertainties (See Appendix \ref{AppendixKlyshkoUncertainty} for details on the estimated uncertainties  $\sigma_{\eta_{i}^{tot}}$). 
These results are consistent with the experimental observations and they would correspond to the APD detection efficiencies in situations with lossless channels and sources, for which $T_{i}=1$, see Eq.~(\ref{model equation}). However, the large discrepancy between the efficiencies from the Klyshko method $\eta^{tot (KM)}_{i}$ and the conventional method  $\eta_{Di}^{Conv}$ are the result of non-unit channel efficiencies $T_{i}$ for each SPDC mode, which is a cumulative effect of source inefficiencies and individual channel losses (See Appendix \ref{Klyshko state losses}). 
Moreover, a reliable implementation of the Klyshko method to accurately determine the APD efficiencies $\eta_{Di}$ requires an accurate determination of these channel efficiencies $T_{i}$.
	
Table \ref{Optical loss} in Appendix \ref{AppendixLoss} lists the losses of the free-space optical channels in our experiment for both optical modes, which are $<1\%$ for each channel. These free-space channel losses do not account for the losses in the SPDC source, which are the largest losses within the system.
Based on the total efficiencies $\eta^{tot(KM)}_{i}$ determined from the Klyshko method and the APD efficiencies  $\eta_{Di}^{Conv}$ from the conventional method, we estimate the efficiency of the SPDC source in each optical mode to be $T_{1} ^{SPDC}= 0.1787$ and $T_{2}^{SPDC} = 0.1722$ for APD1 and APD2, respectively.
This large inefficiency in the SPDC source is mostly due to mode mismatch and coupling of the output modes of the crystal into the optical fiber \cite{bennink2010optimal,ljunggren2005optimal,pereira2013demonstrating,dixon2014heralding}, losses in spectral filtering, and losses in optical components within the source. Other factors contributing to noise and uncorrelated emission include crystal fluorescence, non-optimal crystal temperature for optimal mode matching, and laser mode hops.
All these effects are solely technical limitations in the SPDC source in our experiment, which can be mostly eliminated in new designs of portable bi-photon SPDC sources, optimized for this purpose. 
The precise calibration and optimization of these portable bi-photon SPDC sources will allow for the reliable implementation of the Klyshko method for on-site calibration of SPDs with high accuracy in remote locations, including satellites for earth-to-space quantum communications.

\section{Conclusion} \label{Conclusion}

We investigate the feasibility of the Klyshko method for on-site calibration of single photon detectors (SPDs) with high accuracy, based on a compact, portable quantum correlated bi-photon source based on SPDC. In this method, a photon detection in one mode ideally heralds a single photon in the other mode, which can be used as a standard of radiation at the single photon level for calibrating an SPD in absolute terms. We implement the Klyshko method to calibrate SPD avalanche photodiodes (APDs). We further calibrate the APDs using a transfer standard detector, which provides a benchmark to evaluate the performance of the Klyshko method under similar experimental conditions. We theoretically investigate the impact of the multi-photon character of the SPDC process and system losses in the achievable accuracy for the calibration of SPDs. We observe that multi-photon states increase the probability of photon detection in the detector under test, resulting in an overestimation of its detection efficiency. However, a model of the expected photon counting statistics from the squeezed vacuum in SPDC in principle allows for accurate estimation of the APD efficiencies from the experimental data, assuming that the system losses are known. We conclude that precise calibration and optimization of the SPDC source can enable a reliable implementation of the Klyshko method. Our work provides an understanding of the feasibility of this method for on-site calibration of SPDs without the need for a reference standard, suitable for future realizations of quantum networks with remote nodes containing single-photon quantum receivers.
		
\begin{acknowledgements}
This work was funded by the National Aeronautics and SpaceAdministration (NASA) Award No. 80NSSC32C030 (STTR), theNational Institutes of Health (NIH) Grant No. 1R01GM140284-01,the National Science Foundation (NSF) Grant No. PHY-2210447,and the Department of Energy (DOE) Contract No. CW42943.
\end{acknowledgements}

\section*{Author Declarations}
\subsection*{Conflict of Interest}
The authors have no conflicts to disclose.

\section*{Data Availability}
The data that support the findings of this study are available from the corresponding author upon reasonable request.
			
\appendix

\section{Two-mode squeezed vacuum under loss} \label{Klyshko state losses}

In this appendix, we consider losses in the two-mode squeezed vacuum state $\ket{\psi(r)}$  in Eq. (\ref{SqState}) prior to photon detection, which in general include loss in the source and in the optical channel.  First, we consider losses in one mode of the state $\ket{\psi(r)}$, characterized by the transmittivity $T_1$, followed by photon detection in that mode with non-unit efficiency $\eta_{D1}$. We show that this situation is equivalent to having a total efficiency of $\eta^{tot}_1=T_1\eta_{D1}$ in the conditional post-measurement state \cite{hogg2014efficiencies}. Following a similar procedure, it is then straightforward to integrate losses in the other optical mode of the quantum state $\ket{\psi(r)}$ and photon detection with non-unit efficiency in that mode.

We note that the optical losses can be modeled as the mixing of the input state  $\ket{\psi(r)}$  with a vacuum mode in a beamsplitter $M$ with transmittivity $T_{1}$ \cite{jeffers1993quantum}. Then, tracing over the vacuum port results in the quantum state with loss. This model allows for the incorporation of the losses in the source and the optical channels in the quantum state.

Consider loss in the heralding mode, say $\ket{n}_H$. After the beamsplitter $M$ \cite{leonhardt1997measuring}, the state  $\ket{\psi(r)}$  becomes: 
\begin{equation}
	{\small \begin{split}
			\ket{\psi(r)}_{T_{1}} &= M\ket{\psi(r)} \\
			&=  \sech(r)\sum_{n=0}^{\infty}\sum_{k=0}^{n} \binom{n}{k}^{1/2}t_1^k r_1^{n-k}      \tanh[n](r)\ket{k,n-k}_H\ket{n}_V,
	\end{split}		}
	\label{lossy state}
\end{equation}

where $T_{1} = 1-R_1 = t_1^2 $ is the transmittivity of the beam splitter and $t_1^2 + r_1^2 = 1$. The density matrix of the state $\rho_{T_1}(r)  = 	\ket{\psi(r)}_{T_1}\bra{\psi(r)}_{T_1}$ is:
\\	
\begin{equation}
	\begin{split}
		\rho_{T_1}(r) 
		&= \sech[2](r) \sum_{n_1=0}^{\infty}\sum_{k_1=0}^{n_1}\sum_{n_2=0}^{\infty}\sum_{k_2=0}^{n_2}\tanh[n_1 + n_2](r) \\
		& \times\binom{n_1}{k_1}^{1/2}\binom{n_2}{k_2}^{1/2}  t_1^{k_1} r_1^{n_1 - k_1}  \\
		&  \times \big(  \ket{k_1,n_1-k_1}\bra{k_2,n_2-k_2}  \big)_H \otimes \big(\ket{n_1}\bra{n_2}\big)_V.
	\end{split}
	\label{lossy density matrix}
\end{equation}
The POVM element describing a single photon detection by D1 with detection efficiency $\eta_{D1}$ in this mode is:
\begin{equation}
	\Pi_{ON}^{D_1} = \sum_{n=0}^{\infty}   \bigg( 1- (1-\eta_{D1})^n \bigg) \ket{n}\bra{n}		
	\label{POVM_ON_helralding}
\end{equation} 

Applying $\Pi_{ON}^{D_1}$ to $\rho_{T_1}(r)$, the conditional (unnormalized) state $\tilde{\rho}_{B|D1}^{T_1}$ in the heralded mode becomes:
\\
\begin{equation}
	\begin{split}
		\tilde{\rho}_{B|D1}^{T_1} &= Tr_{A}\bigl\{\Pi_{ON}^{D1}\rho_{T_1}(r) \bigr\} \\
		&= \sech[2](r)\sum_{n=0}^{\infty}\sum_{k=0}^{n} \tanh[2n](r) \binom{n}{k} t_1^{2k} r_1^{2(n-k)} \\
		&\times \big( 1- (1-\eta_{D1})^k \big)  \big( \ket{n}\bra{n} \big)_V\\
		&= \sech[2](r)\sum_{n=0}^{\infty} \tanh[2n](r) \sum_{k=0}^{n} \binom{n}{k} \\
		&\times\bigg( T_1^{k} R_1^{(n-k)} - T_1^{k}(1-\eta_{D1})^k R_1^{(n-k)} \bigg) 
		\big( \ket{n}\bra{n} \big)_V \\
		&= \sech[2](r)\sum_{n=0}^{\infty} \tanh[2n](r) \\
		&\times \big[  (T_1+R_1)^n - (T_1(1-\eta_{D1}) + R_1)^n \big]\big( \ket{n}\bra{n} \big)_V \\
		&= \sech[2](r)\sum_{n=0}^{\infty} \tanh[2n](r)  \big[ 1-(1-\eta_{D1} T_1)^n \big] \big( \ket{n}\bra{n} \big)_V\\
		&= (1-\zeta) \sum_{n=0}^{\infty} \zeta^n  \big[ 1-(1-\eta_{D1} T_1)^n \big] \big( \ket{n}\bra{n} \big)_V\\
		\tilde{\rho}_{B|D1}^{T_1} &= \rho_{th}(\zeta) - \frac{1-\zeta}{1 - \zeta(1-\eta_{D1} T_1)}\rho_{th}(\zeta(1-\eta_{D1} T_1)),
	\end{split}
	\label{unnormalized lossy dm 2}		
\end{equation}

where $\rho_{th}(\zeta)$ is the thermal state (Eq. (\ref{thermal state})).\\
\\
Eq. (\ref{unnormalized lossy dm 2}) shows that linear loss in the state, characterized by a transmittivity $T_1$, followed by a photon detection measurement $\Pi_{ON}^{D_1}$ with efficiency $\eta_{D1} $ is equivalent to having a detector with an overall detection efficiency $\eta^{tot}_{1} = \eta_{D1} T_1$.\\
Normalizing Eq. (\ref{unnormalized lossy dm 2}), the conditional state $ \rho_{B|D1}^{T_1}$ becomes:
\\
\begin{equation}		
	\begin{split}
		\rho_{B|D1}^{T_1} &=  \frac{1 - \zeta(1-\eta^{tot}_{1})}{\zeta \eta^{tot}_{1}} \\
		&\times \bigg[\rho_{th}(\zeta) - \frac{1-\zeta}{1 - \zeta(1-\eta^{tot}_{1})}\rho_{th}(\zeta(1-\eta^{tot}_{1}))\bigg].
	\end{split}
\end{equation}

Following the procedure delineated above, it is then straightforward to consider linear losses in the heralded (second) mode ($\ket{n}_V$) of $\ket{\psi(r)}$ by considering a beam splitter with transmittivity $T_2$ followed by photon detection by D2 with detector of efficiency $\eta_{D2}$. In the same way as for the first mode, the channel efficiency ($T_2$) and the detection efficiency ($\eta_{D2}$) for the second mode can be combined into a single parameter as the overall detection efficiency $\eta^{tot}_{2} = \eta_{D2} T_2$ for the heralded mode.\\ 
As a result, the probability $P_{D2|D1}^{ON} $ of a photon detection event in D2 conditioned on having observed a photon in D1 is given by:

\begin{equation}
	\begin{split}
		P_{D2|D1}^{ON} &=  Tr\bigl\{\Pi_{ON}^{D2}\rho_{B|D1}^{T_1}\bigr\} \\
		&=\biggl\{1-\left(\frac{1-\zeta(1-\eta^{tot}_{1})}{\zeta\eta^{tot}_{1}}\right)
		(1-\zeta)\\
		&~~\times\left[\frac{1}{1-\zeta(1-\eta^{tot}_{2})}
		-\frac{1}{1-\zeta(1-\eta^{tot}_{2})(1-\eta^{tot}_{1})}\right]\biggr\},
	\end{split}
\end{equation}

which is Eq. (\ref{model equation}) in the main manuscript.		

\section{Accidental coincidences in the experiment}\label{accidental appendix}

\begin{figure}[b]
	\centering
	\includegraphics[scale = 1.0]{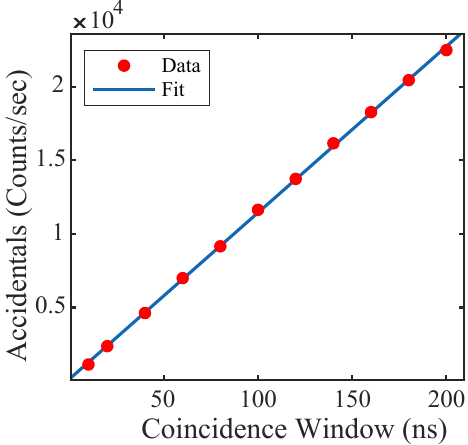}
	\caption{Accidental coincidences as a function of coincidence window $CW$. The width ($W_{D2}$) of the digital signal inside FPGA given a detection in (heralded) detector D2 is set to 20 ns in this case.}
	\label{fig: accidental plot}
\end{figure}

The accurate determination of the effective coincidences requires information about the rate of accidental coincidences due to uncorrelated photons from the source, dark counts, background, etc. We determine this rate by studying the coincidences with a relative delay of 300 ns between the heralding and heralded detectors, which is much longer than the coherence time of the bi-photon state. The rate of accidental coincidences depends on the effective coincidence window in our experiment $CW_{eff}$, which we model as:

\begin{equation}
	CW_{eff} = CW + W_{D2} - \Delta W.
	\label{coin model}
\end{equation}
Here $CW$ is the coincidence window set in the counting electronics based on the FPGA, conditioned to a detection in the heralding detector D1.
$W_{D2}$ is the width of the digital signal inside the FPGA corresponding to a detection event in the (heralded) detector D2. We set $W_{D2}=20$ ns. Finally, $\Delta W$ is the resolution \cite{beck2007comparing,spiess2023clock} of our coincidence electronics. Given the FPGA master clock of 100 MHz, we expect $\Delta W\approx10$ ns. 

Figure \ref{fig: accidental plot} shows the observed accidental coincidences as a function of coincidence window $CW$ over one second. We observe a linear relation between the accidentals and $CW$, as expected. The blue line is a linear fit to the data, where the independent variable corresponds to $CW$  and the slope corresponds to the rate of accidentals. We observe a very good agreement between the experimental observations and the linear fit. From this fit, we determine that $\Delta W\approx12$ ns, which is consistent with the expected resolution of the coincidence counting program in the FPGA. This study shows that the accidental coincidences come from random, uncorrelated events and allows us to accurately determine the accidental coincidences for a given coincidence window chosen in the experiment.

\section{Uncertainty Budget in measurement of detection efficiencies} \label{AppendixLoss}
		
The uncertainty of the estimation of the detection efficiency of the SPDs depends on different factors involved in the calibration procedure. Table I and Table II describe these factors involved in the calibrations based on the conventional and the Klyshko methods, respectively, and their contributions to the uncertainty in the determination of the APD detection efficiencies.
		
\begin{table}[h]
	\centering
	\caption{Uncertainty estimates in the conventional calibration method for the two detectors.}
	\begin{tabular}[t]{ccc}
		\hline
		&Value&Standard Deviation\\
		\hline
		Attenuation&$3.51\times10^{-6}$&$1.98\times10^{-9}$\\
		Trap detector (V)&-4.51&9.24$\times10^{-4}$\\
		Incident photon count rate\\ (Counts/sec)&1$\times10^{5}$&1.8$\times10^{-2}$\\
		Perkin Elmer detected\\ count rate
		(Counts/sec)&$63.7\times10^4$&3.06$\times10^2$\\Count detected count rate\\(Counts/sec)&57.5$\times10^4$&2.3$\times10^2$\\ 
		$\eta^C_{D1}$ (Perkin Elmer)&0.637&$3.06\times10^{-3}$\\
		$\eta^C_{D2}$ (Count)&0.575&$2.3\times10^{-3}$\\
		\hline
	\end{tabular}
	\label{Conv Uncertainty}
\end{table}%
		
\begin{table}[h]
	\centering
	\caption{Optical channel loss estimates for the two optical channels for APD1 and APD2 in the Klyshko method.}
	\begin{tabular}[t]{cccc}
		\hline
		&Trap detector (V)&Standard Deviation&Loss(\%)\\
		\hline
		Incident light&-3.136&$1.8\times10^{-3}$&-\\
		Channel 1&-3.120&$6.4\times10^{-3}$&0.5\\
		Channel 2&-3.112&$3.2\times10^{-3}$&0.77
	\end{tabular}
	\label{Optical loss}
\end{table}		
		
\section{Uncertainty from the Klyshko method} \label{AppendixKlyshkoUncertainty}
		
For the portable, commercial SPDC source in our experiment, there is a small range of pump powers over which we can reliably implement the Klyshko method. In this range, the pump laser is mode-hop free, and the source yields low enough photon generation rates to minimize, and practically avoid, accidental coincidences from uncorrelated SPDC process and aferpulsing effects. For this range of pump powers, the experimental data in Figure \ref{fig:de vs dac} in the main text shows a near linear behavior, which results in the theoretical model in Eq. (\ref{model equation}), which is a nonlinear model, overfitting the data. As a result, the uncertainties in the estimates $\eta_{i}^{tot(KM)}$ from the fit are smaller than the statistical uncertainties from the data. Therefore, the uncertainty in $\eta_{i}^{tot(KM)}$ is dominated by statistical uncertainties.

We obtain the statistical uncertainty of $\eta^{tot (KM)}_{i}$ from the conditional probabilities in Eq.~(\ref{model equation}) using error propagation. Since $P_{D2|D1}^{ON}(\eta^{tot}_{1},\eta^{tot}_{2},r)$ depends on three independent parameters $\eta^{tot}_{1}$, $\eta^{tot}_{2}$ and $r$, the uncertainty in the estimation of the detection efficiency for the heralded detector D2 ($\eta^{tot}_{2}$ in the model) is:
	
\begin{equation}
\sigma_{\eta^{tot}_{2}} = \sqrt{\bigg(\frac{\mathrm{d}\eta^{tot}_{2}}{\mathrm{d}P_{D2|D1}^{ON}}\bigg)^2	\sigma_{P_{D2|D1}^{ON}}^2	},
\end{equation}

where $\sigma_{P_{D2|D1}^{ON}}$ are the uncertainties in the conditional probability $P_{D2|D1}^{ON}$ observed in the experiment. From implicit differentiation of Eq.~(\ref{model equation}), we obtain:

{\small 		\begin{equation}
	\begin{split}
		\frac{\mathrm{d}\eta^{tot}_{2}}{\mathrm{d}P_{D2|D1}^{ON}} &=  \left(\frac{\eta^{tot}_{1}}{1-\zeta(1-\eta^{tot}_{1})}\right)
		\frac{1}{(1-\zeta)}\\
		&~\times\left[\frac{1}{\big[1-\zeta(1-\eta^{tot}_{2})\big]^2}
		-\frac{1-\eta^{tot}_{1}}{\big[1-\zeta(1-\eta^{tot}_{2})(1-\eta^{tot}_{1})\big]^2}\right]^{-1}.
	\end{split}
	\label{derivative error}			
\end{equation}}
		
\begin{figure}[t]
	\centering
	\includegraphics[scale = 1.0]{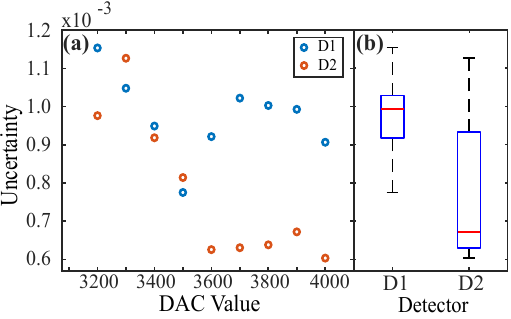}
	\caption{(a) Estimated uncertainties  $\sigma_{\eta^{tot}_{1}}$ and  $\sigma_{\eta^{tot}_{2}}$ for the efficiencies $\eta^{tot}_{1}$ and $\eta^{tot}_{2}$, respectively, evaluated at different pump powers corresponding to the data points shown in Figure \ref{fig:de vs dac}.(b) Boxplot representation of the variation in the estimated uncertainties around the median (red line) for the calibrated detectors. }
	\label{fig: propagated error}
\end{figure}

We note that $\sigma_{\eta^{tot}_{2}}$ depends on the value of the parameters $\eta^{tot}_{1}$, $\eta^{tot}_{2}$, and $r$. For these values, we use the best estimates of these parameters obtained from the fit of the data in Figure \ref{fig:de vs dac}. We then evaluate $\sigma_{\eta^{tot}_{2}}$ at the specific pump powers corresponding to the data points shown in Figure \ref{fig:de vs dac}.
A similar procedure can be followed to obtain the uncertainty  $\sigma_{\eta^{tot}_{1}}$ for the efficiency of detector D1 from the conditional probability $P_{D1|D2}^{ON}$.  Figure \ref{fig: propagated error}(a) shows the uncertainties  $\sigma_{\eta^{tot}_{1}}$ and  $\sigma_{\eta^{tot}_{2}}$ for the efficiencies $\eta^{tot}_{1}$ and $\eta^{tot}_{2}$, respectively, for pump powers 
corresponding to the data points shown in Figure \ref{fig:de vs dac}. Figure \ref{fig: propagated error}(b) shows the boxplot representing the distributions of the estimated uncertainties and their skewness. The median of the uncertainties (red lines in the boxplot) are $0.99 \times 10^{-3}$ for D1 and $0.67 \times 10^{-3}$ for D2, respectively.		
		
\section*{References}

\bibliography{Bibliography.bib} 

\end{document}